\documentclass[review]{elsarticle}

\usepackage{color}
\usepackage{array}
\usepackage{graphicx}
\usepackage{siunitx}
\usepackage{booktabs}
\usepackage{epstopdf}
\usepackage{amsmath}
\usepackage{siunitx}
\usepackage{multirow}
\usepackage{multicol}
\usepackage{longtable}

\usepackage{setspace}
\usepackage{psfrag}     

\usepackage{lineno,hyperref}
\modulolinenumbers[1]

\journal{Energy Conversion and Management}

\begin{document}
\begin{spacing}{2.0}  

\begin{frontmatter}

\title{Adaptive operation strategy for voltage stability enhancement in active DMFCs}


\author[Address1]{Qin-Wen YANG}
\author[Address1]{Xu-Qu HU\corref{Corespond}}
\cortext[Corespond]{Corresponding author}
\ead{huxuqu@gmail.com}
\author[Address1]{Xiu-Cheng LEI}
\author[Address1]{Ying ZHU}
\author[Address1]{Xing-Yi WANG}
\author[Address2]{Sheng-Cheng JI}

\address[Address1]{State Key Laboratory of Advanced Design and Manufacturing for Vehicle Body, College of Mechanical and Vehicle Engineering, Hunan University, Changsha, 410082, China}
\address[Address2]{Beijing Aeronautical Science \& Technology Research Institute of COMAC, Beijing, 102211, China}

\begin{abstract}
An adaptive operation strategy for on-demand control of active direct methanol fuel cells (DMFCs) is proposed as an alternative method to enhance the voltage stability.
A simplified semi-empirical model is firstly developed to describe I-V relationships based on uniform-designed experiments.
It is then embedded into multi-objective optimizations to construct the adaptive operation strategy. 
Experimental studies are implemented on different DMFC systems to validate the proposed semi-empirical model, control strategy and system response to operational adjustments.
Numerical simulations are also performed to investigate the underlying mechanisms of the proposed adaptive operation strategy.
The results show that the adaptive operation strategy provides possibilities for voltage stability enhancement without the sacrifice of energy conversion efficiency.
The adaptive operations are also found to be able to extend the range of operating current density or to decrease the voltage deviation according to one’s requirements.
Moreover, the response of DMFCs to operational adjustments is quick, which further validates the effectiveness and feasibility of the adaptive operation strategy in practical applications.
The proposed strategy contributes to a guideline for the better control of output voltage from operating DMFC systems.
\end{abstract}


\end{frontmatter}


\section{Introduction}

Direct methanol fuel cell (DMFC) has emerged in recent years as a promising power source for various kinds of electronic devices \cite{Chau1999ECM, Ji2017ECM}. It can provide high energy density and almost instant recharging with simple system configurations \cite{Kamaruddin2013RSER,Taner2018energy}. 
But to move towards real-time operations of DMFCs that meet a commercially acceptable standard, further improvements are still needed especially in the fields of systematic stability \cite{Taner2015APE, Zhou2016IEEE, ZHOU2017IEEE, wu2017realtimeIEEE}. 
During the past decades, lots of research efforts have been dedicated to enhance the system stability of fuel cells by improving the catalyst loading \cite{karim2013CatalystAE, Patel2015JPS}, the fuel cell structure \cite{fukuhara2014SSI, Wang2017ECM, Ozden2017IJHE, Wilberforce2017IJHE}, the fuel delivery system \cite{Kamaruddin2013RSER, Mehmood2015JPS} and the water management \cite{Taner2015, taner2017micro}, etc.

One problem that received relatively less attention is the voltage instability. 
Because of the over-potential effects, the output voltage from DMFCs is inevitably unsteady as the current density is often required to change in a wide range during practical applications.
An obvious voltage deviation is, therefore, frequently encountered even if the DMFC was assumed to be operated in the theoretical ohmic loss range \cite{Mehmood2015JPS, Barbir2012PEM}. Many published experimental results have shown that the output voltage could be very sensitive to a small variation of current density \cite{Patel2015JPS, Ozden2017IJHE, WANG2011reviewAE}. 
Although the voltage stabilizers can be helpful to a certain extent, an operation strategy is also in great need to avoid additional energy loss caused by voltage stabilizations. 
Moreover, it can be necessary while the voltage deviation is beyond the input limit of voltage stabilizers.

Several control strategies have been applied to analyze and enhance the system stability of DMFCs \cite{Bizon2014ECM, Bizon2015ECM}, such as PID control algorithms \cite{13Zenith2010JPC}, neural network methods \cite{16Chang2012IJICIC}, model predictive control schemes \cite{17Fan2013JESTR, 18Keller2017CEP}, fuzzy logic \cite{Zhou2017JPS} and extremum seeking controls \cite{Zhou2017ECM}.
However, most of them were designed to work under the assumption of small disturbance (or little noise).
Since the operating conditions in commercial electric equipments would change sharply, the influence of large disturbance on the voltage stability needs to be taken into account. 
An adaptive operation strategy that is suitable for considering large disturbance in complex operating conditions becomes the initial motivation of the present study.

However, it is full of challenges, as it involves accurate modeling of DMFC systems to study the underlying mechanisms of multiple operating parameters.
During the past decades, several mechanical models were proposed to study the energy conversion process of DMFCs, from one-dimensional \cite{19Oliveira2008IJHE, 20Ko2010Energy}, two-dimensional \cite{ZHOU2017IEEE, 22Birgersson2003JES, 23Yan2008IJHMT, zhou2018ECM} to three-dimensional \cite{26Yu2013IJHE, Heidary2016ECM} point of view.
Considering about large variations of operating parameters, we have proposed an effective semi-empirical model to describe I-V relationship and developed a three-dimensional Computational Fluid Dynamics (CFD) model to study the DMFC performance \cite{29Yang2011JPS}. More recently, it has been successfully applied to determine the optimal operating conditions in DMFC systems \cite{30Hu2017Energies}.
Although the effects of operating conditions (such as temperature, methanol concentration and solution/air flow rates) on the DMFC performance can be well estimated using those existing models \cite{27Chu2006EA, 28Silva2012AMC, zhou2016IEEEEC}, an effective operation strategy can still be necessary for determine appropriate operational adjustments to enhance voltage stability in practical applications.

In the present study, an adaptive operation strategy is developed. 
Its performance and underlying mechanisms have also been systematically studied using analytical, numerical and experimental techniques. 
The strategy establishment is presented in details in Section \ref{Sec:MethodMods}.
The theoretical, experimental and numerical studies about the necessity, feasibility and superiority of the proposed strategy are organized in Section \ref{Sec:Res}.
Experimental validations about the semi-empirical model, control strategy and system response are provided in Section \ref{Sec:ExpValid}.
Conclusions and perspectives were summarized in Section \ref{Sec:ConPers}.

\section{Methods and models}
\label{Sec:MethodMods}

The adaptive operation strategy is developed based on the integration of experimental, analytical and numerical techniques.
Uniform-designed experiments are firstly performed to study the I-V relationships under various groups of operating parameters.
A simplified semi-empirical model is then developed to quantitatively describe DMFC performances in experiments.
Based on the semi-empirical model, the adaptive operation strategy is developed using multi-objective optimizations.
The underlying mechanism of adaptive DMFC operations are also studied using numerical simulations.

\subsection{Experimental setup and design}
\label{ExpSetup}

\begin{figure}[htbp]
\centering
\includegraphics[width=0.875\linewidth]{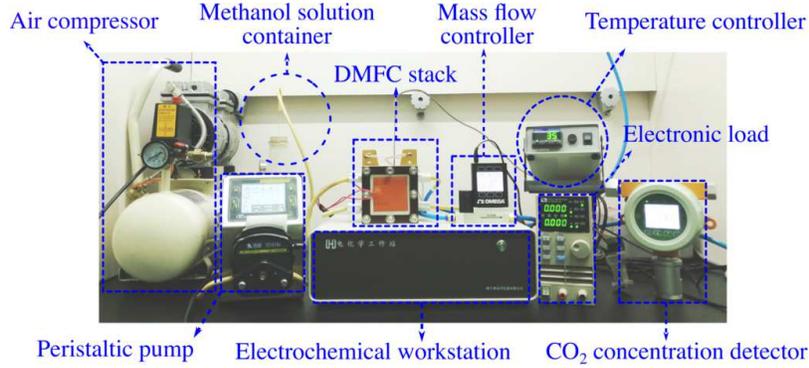}
\caption{Experimental platform of a single-cell DMFC system.}
\label{fig:2}
\end{figure}

Fig. \ref{fig:2} shows the experimental platform applied in the present study.
A single-cell DMFC stack (TeskStak, Parker Hannifin Energy Systems) consists of a five-layer Membrane Electrode Assembly (MEA) sandwiched by two graphite end plates (for anode and cathode).
Serpentine flow channel of 12 U-turns is integrated inside the end plates, which takes a rectangular cross-section (30.94 $\times$ 30.65 $\rm mm^2$) and a total length of 415.06 $\rm mm$. 
At the anode side, the methanol solution composed by deionized water and pure methanol is flowed into the channel by a peristaltic pump (BT300LC). 
At the cathode side, an air compressor (OUTSTANDING OTS-550) regulated by a rotameter (OMEGA FMA-2605A) is used to pump the air into the fuel cell.
The working temperature of the DMFC stack is maintained by a supplementary heating apparatus that is controlled by a temperature controller (Omega CSC32). 
An electronic load device (ITECH it8211) is used to change the current density to different levels and to measure the corresponding values of the voltage.
The DMFC performance is monitored by an electrochemical workstation (CHI660E), while the production of $\rm CO_2$ is measured by a $\rm CO_2$ concentration detector (JA500-CO2-IR1).

\begin{table}[thbp]
\centering
\caption{Value ranges of main operating parameters.}
\label{tab:expcase}
\begin{tabular}{>{\small}c >{\small}c >{\small}c >{\small}c}
\hline
Name & Value range & Name & Value range \\
\hline
 Temperature ($\rm{K}$) & [298, 343]   &  Methanol concentration ($\rm{M}$)  & [0.25, 2]  \\
 Air flow rate ($\rm{ccm}$) & [81.2, 186]  &  Feed solution flow rate ($\rm{ccm}$) & [3.5, 5.5] \\ 
\hline
\end{tabular}
\end{table}

Four main operating parameters are considered in experimental study, including the working temperature ($T$), the methanol concentration ($C_M$), the input flow rates of feed methanol solution ($F_M$) and air ($F_{A}$). 
The value ranges of operating parameters are determined according to literature review and practical experience, as summarized in Table \ref{tab:expcase}.
Based on the uniform design principal, a systematic experimental work is designed to study the influence of operating parameters \cite{Fang2000Tech}.
As five levels are selected for each operating parameters, 65 experimental tests are designed and performed. 
Among them, 45 cases are decided by uniform design, and the other 20 ones are supplementary for the analysis of DMFC performance degradation.
Details about the operating parameters for these uniform-designed experimental cases are summarized in Appendix, Table \ref{tab:expconditions}. 

\subsection{Semi-empirical model}
\label{Sec:semi-empiriMod}

Since the degradation of DMFC performance was observed in experiments, a linear regression method is applied in the collection process of experimental data \cite{29Yang2011JPS, Kianimanesh2013FC}.
Careful investigation is implemented to analyze the compensated experimental data, and a semi-empirical model can be deduced to describe the relationship between the operating parameters and the output voltage.

A semi-empirical model was successfully developed to quantify the influence of main operating parameters on the DMFC output in our previous studies \cite{26Yu2013IJHE, 29Yang2011JPS}. 
However, it contains a large number of undetermined coefficients which requires considerable computation cost for identification. 
A simplified semi-empirical model with less coefficients is then proposed in the present study, which can be expressed as,

\begin{align}
\centering 
V_{cell} & = V_{th} + k_1 T + k_2 T ln(C_{M}) + k_3 T ln(F_{A}) + k_4 \notag \\
& - k_5e^{(k_6/T+k_7)}j  -  (k_8 j^2 + k_9T + k_{10}C_{M}^2+k_{11}C_{M} + k_{12}) \notag \\ 
& \times \left [ ln(j) + k_{13} + k_{14} ln \left( \frac{1}{ C_{M} - \frac{j}{k_{15}e^{(-k_{16}/T)}C_{M}} } \right)  \right ]  \\
& - (k_{17}j + k_{18}T + k_{19}C_{M}^2 + k_{20}C_{M} + k_{21}) \notag \\
& \times [ln(j) + k_{22}ln(F_{A}) + k_{23} ] + k_{24} j^2 ln(F_{M}), \notag
\label{Eq:2}
\end{align}
where $V_{cell}$ and $j$ are the output voltage and the current density, respectively. $V_{th}$ denotes the reversible 'no-loss' voltage which theoretically equals to 1.21 V. 
$T$, $C_M$, $F_M$ and $F_{A}$ denote the operating parameters as mentioned above, and $k_i~(i=1, 2, ..., 24)$ denotes different undetermined coefficient.
Generally speaking, this model is based on electrochemical theory and constructed through integration of three sub-models, i.e., the open circuit sub-model ($k_1$-$k_4$), the resistance sub-model  ($k_5$-$k_7$) and the closed circuit one ($k_8$-$k_{24}$). 

For coefficient identifications, the reference data can be collected from more than 65 experimental tests based on the principal of uniform design (as mentioned in Section \ref{ExpSetup}).
Firstly, the coefficients in resistance sub-model are obtained through regression of experimental data of area-specific resistance varying with temperatures.
The coefficients in open circuit sub-model can be then determined by regression of experimental results of open circuit voltage changing with operating parameters.
Finally, the rest coefficients in closed circuit sub-model are identified through nonlinear regression of experimental I-V curves.
The determined coefficients are listed as follows, $k_1=-3.7534 \times 10^{-5}$, $k_2=-3.1534 \times 10^{-4}$, $k_3=6.6200 \times 10^{-5}$, $k_4=-0.7499$, $k_5=6.9897$, $k_6=916.91$, $k_7=-4.6392$, $k_8=-0.8801$, $k_9=-0.5791$, $k_{10}=-4.8053$, $k_{11}=4.8053$, $k_{12}=-1.2135$, $k_{13}=-36.4865$, $k_{14}=2.8580 \times 10^{-4}$, $k_{15}=5.34657 \times 10^7$, $k_{16}=5182.4$, $k_{17}=-0.1160$, $k_{18}=0.5793$, $k_{19}=4.8062$, $k_{20}=-4.8062$, $k_{21}=1.2016$, $k_{22}=-8.2586 \times 10^{-4}$, $k_{23}=-36.4679$ and $k_{24}=29.8714$.

\begin{figure}[t]
\centering
\includegraphics[width=0.875\linewidth]{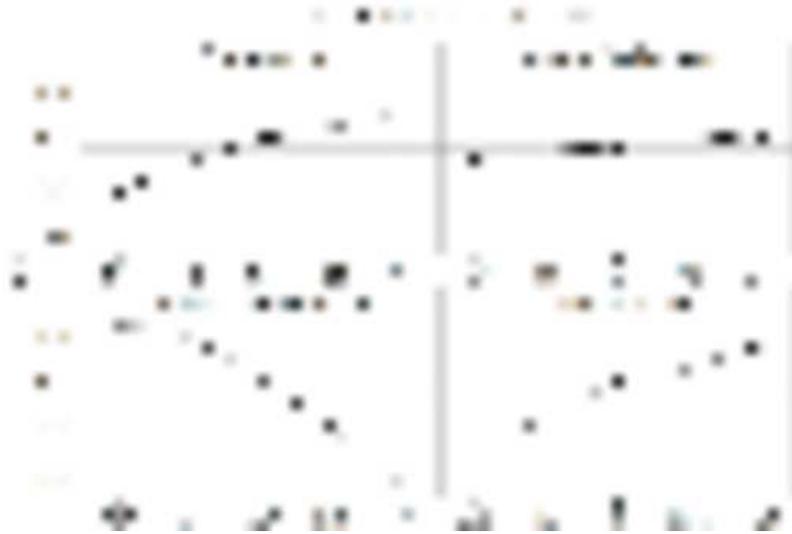}
\caption{Contributions of main operating parameters to output voltage.}
\label{fig:FigParamaticAnalys}
\end{figure}

A parameter sensitivity analysis about the quantitative effects of the main operating parameters on the output voltage is performed based on the statistics package Minitab (Fig. \ref{fig:FigParamaticAnalys}).
It shows that the output voltage is sensitive to all the four main operating parameters, even if the change of methanol flow rate seems to take relatively small effect compared with the other three parameters. 
A systematic trend is observed for all the four parameters, with the voltage deviations from its mean 
trending largest at the extreme ranges of those parameters. 
Moreover, the increase of methanol concentration is observed to be accompanied by a decrease of output voltage, but the situation for the other three parameters reverses. 
This quantitative analysis could provide a parameter control guideline for engineers, which could also be highly beneficial for the design of our operation strategy.

\subsection{Adaptive operation strategy}

\begin{figure}[htpb]
\centering
\includegraphics[width=1.0\linewidth]{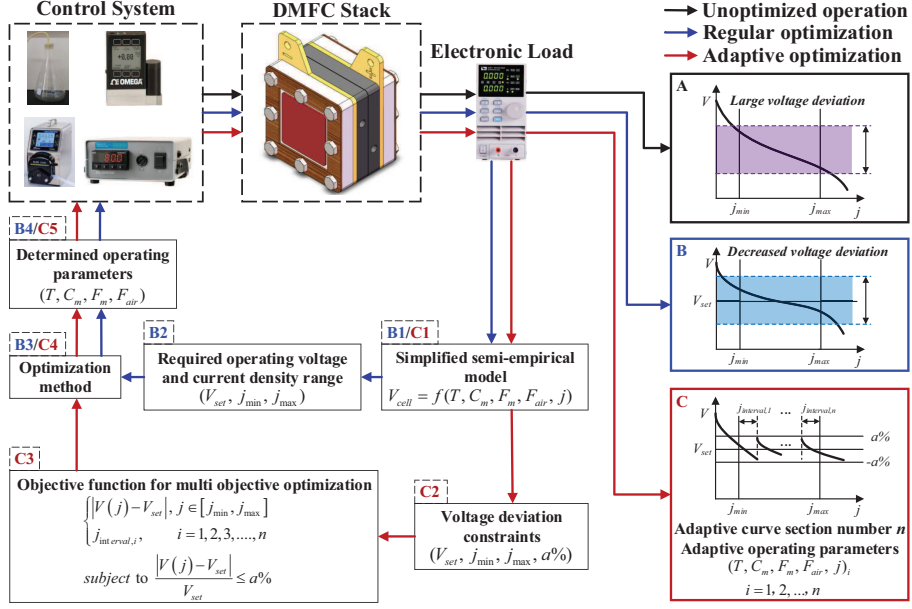}
\caption{Schematic of operation strategies for DMFC systems.}
\label{fig:flowchartOPT}
\end{figure}

A commercial power source is usually required to provide a steady voltage output $V_{set}$, while its operating current density varies in a range of $[j_{min}, j_{max}]$.
However, the problem of large voltage deviation is frequently encountered in unoptimized DMFC operations (Fig. \ref{fig:flowchartOPT} (A)).
It would worsen the performance of control systems and increase the difficulty for practical uses of DMFC systems.
The enhancement of voltage stability around $V_{set}$ with current density varying in $[j_{min}, j_{max}]$ is thus one of the critical issues for the commercialization of DMFC systems. 
Optimized operation strategies are then proposed for such purpose in this section. 
It is of importance to notice that the following optimizations are all based on our simplified semi-empirical model (see Section \ref{Sec:semi-empiriMod}), as it provides a fast and reliable way to determine the output voltage from a specific current density $j$ and the corresponding operating parameters.

A regular optimization strategy (Fig. \ref{fig:flowchartOPT} (B)) can be deduced by direct utilizations of classical optimization methods, such as genetic algorithm (GA) or particle swarm optimization (PSO). 
More specifically, the classical optimization method is used to minimize the voltage deviation $|V_{cell}-V_{set}|$ in the full range of required current density $[j_{min}, j_{max}]$ and determine the specific group of operating parameters $[T,~C_M,~F_M,~F_{A}]$, as shown in Fig. \ref{fig:flowchartOPT} (B1-B4).
Compared with unoptimized cases, the regular optimization could decrease the voltage deviation to a certain degree. 
However, it can only provide one single I-V curve, for which the voltage stability could still not guarantee the relatively small voltage deviation $\pm a\%$ permitted in practical use.
As the I-V curves can be switched by adjusting operating parameters, an appropriate operational adjustment can be highly beneficial for the enhancement of voltage stability in DMFC systems.

Inspired by the logical deductions as mentioned above, an adaptive operation strategy is further proposed as demonstrated in Fig. \ref{fig:flowchartOPT} (C).
Its main idea lies in the determinations of appropriate adjustments of operating parameters, to fulfill the required small voltage deviation $\pm a\%$ and the large range of working current density $[j_{min}, j_{max}]$ (Fig. \ref{fig:flowchartOPT} (C1-C2)).
Such a real-time adjustment brings out a switch of I-V curves, the modified DMFC performance is then described by an integration of serval sections from different I-V curves.
The determinations of each I-V curve section involve two objectives, which are the minimization of voltage deviation $|V_{cell}-V_{set}|$ and the maximization of each current density interval $j_{interval}$.
A multi-objective optimization model is then proposed (Fig. \ref{fig:flowchartOPT} (C3)), which can be expressed as

\begin{align}
\centering 
\min f(x,j_{interval}) & =\min \{w_1[V(j_0,x)-V(j_0+j_{interal},x)]+w_2\frac{1}{j_{interval}} \},\notag \\
s.t. & \quad \left\{
\begin{aligned}
& \frac{|V(j_0,x)-V_{set}|}{V_{set}} \leq a\%, \\
& \frac{|V(j_0+j_{interval},x)-V_{set}|}{V_{set}} \leq a\%,
\end{aligned}
\right. \\
for & \quad j \in [j_{min}, j_{max}], \notag
\label{Eq:MultiOPT}
\end{align}
where $j_0$ is the starting current density for each section, and $x$ denotes a random series of operating parameters (i.e., $[T,~C_M,~F_M,~F_{A}]$).
The main multi-objective function $f$ is designed to be composed by two parts, one 'voltage-deviation' subfunction $|V_{j_0}-V_{j_{0}+j_{interval}}|$ measures the voltage variation for a specific I-V curve section, and the other 'interval' subfunction $\frac{1}{j_{interval}}$ accounts for the effect of operating range of current density of each I-V curve sections.
The weighting factors for each subfunctions, $w_1$ and $w_2$, can be adapted to one's specific requirements.

During program implementations, the constraint of permitted voltage deviation $\pm a\%$ is guaranteed inside the 'interval' subfunction. Specifically speaking, once the voltage deviation of certain I-V curve section exceeds the constraint, a switch of operating parameters $x$ is designed to execute  automatically, until the constraint is fully satisfied in the whole range of working current density $[j_{min}, j_{max}]$. 
Several classical optimization algorithms, such as GA, PSO and simulated annealing (SA), have been successfully applied for solving multi-objective optimization problems in civil engineering \cite{asadi2014multi, delgarm2016multi}, solar energy and hydraulic energy systems \cite{wang2015multi, marques2015multi}, etc.
With the help of classical optimization methods, our multi-objective optimization model can be solved to determine the series of operating parameters $x$ and the current density intervals $j_{interval}$ (Fig. \ref{fig:flowchartOPT} (C4-C5)). 
The adaptive operation strategy with appropriate operational adjustments is thus developed.

\subsection{Numerical model}

A three-dimensional numerical model was developed and validated to experimental results in our previous studies \cite{29Yang2011JPS, 30Hu2017Energies}. 
It was designed by integrating the governing equations of continuity, momentum conservation, species transport and electrochemical phenomena. 
It has been successfully applied to investigate the energy conversion process in a DMFC system \cite{30Hu2017Energies}, which is the same one as that we used in the present study. 
Therefore, this well-developed numerical model can be adopted in the present study to investigate the underlying mechanisms of the adaptive operation strategy. 
For simplicity and concision, more details about the numerical model were not contained here, which can be available in Ref. \cite{29Yang2011JPS, 30Hu2017Energies}.

\section{Results and discussions}
\label{Sec:Res}

\subsection{Large voltage variation in unoptimized operations}

\begin{table}[htpb]
\centering
\caption{Operating parameters for testing cases with large voltage variations.}
\label{tab:testcase}
\begin{tabular}{>{\small}c >{\small}c >{\small}c >{\small}c >{\small}c}
\hline
$Case$ & $T$ (\rm{K}) & $C_M (\rm{M})$ & $F_M$ (\rm{ccm}) & $F_A$ (\rm{ccm}) \\
\hline
1 & 323 & 0.25 & 4 & 81.2 \\
2 & 323 & 0.50 & 5 & 140.8 \\
\hline
\end{tabular}
\end{table}

\begin{figure}[t]
\centering
\psfrag{Case 1}[c][c][0.675][0]{Case 1}
\psfrag{Case 2}[c][c][0.675][0]{Case 2}
\psfrag{Regular}[c][c][0.675][0]{Regular}
\psfrag{Adaptive (GA)}[c][c][0.675][0]{Adaptive (GA)}
\psfrag{Adaptive (PSO)}[c][c][0.675][0]{Adaptive (PSO)}
\includegraphics[width=0.7\linewidth]{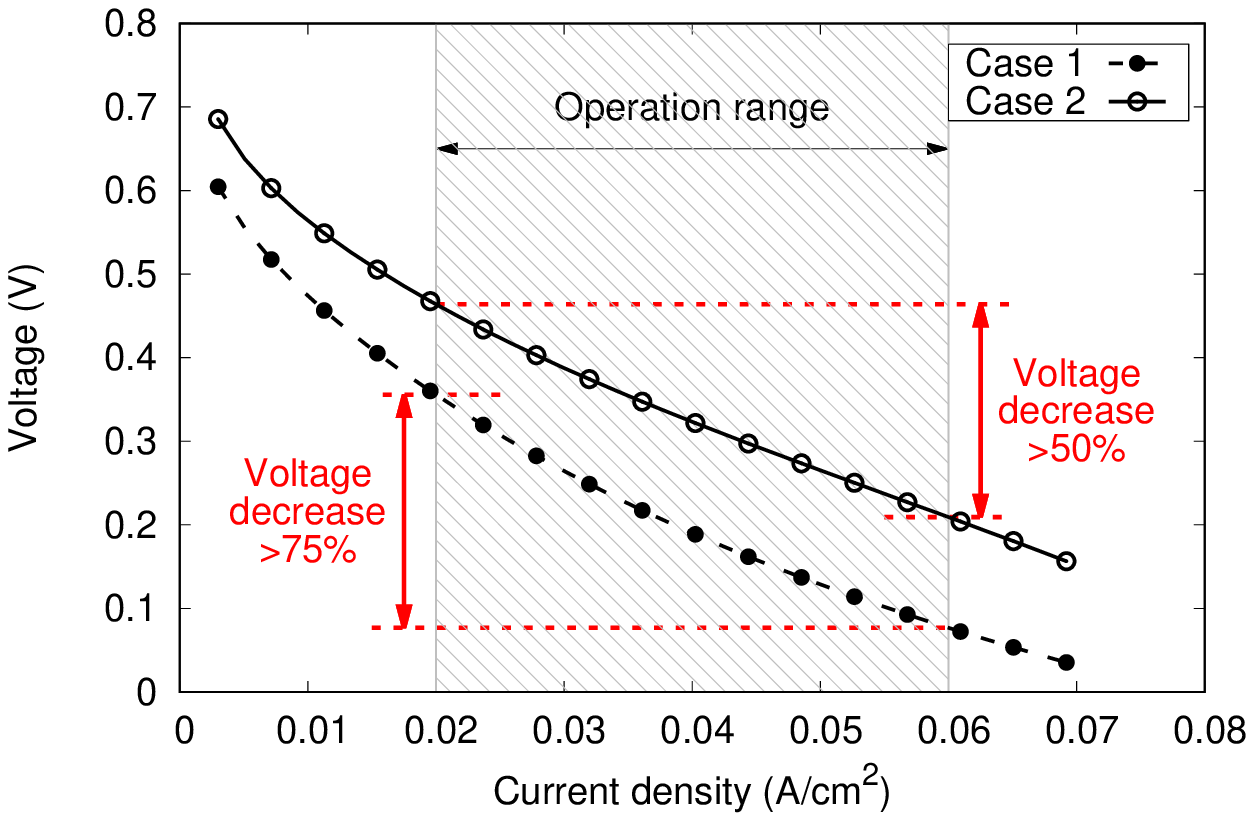}
\includegraphics[width=0.7\linewidth]{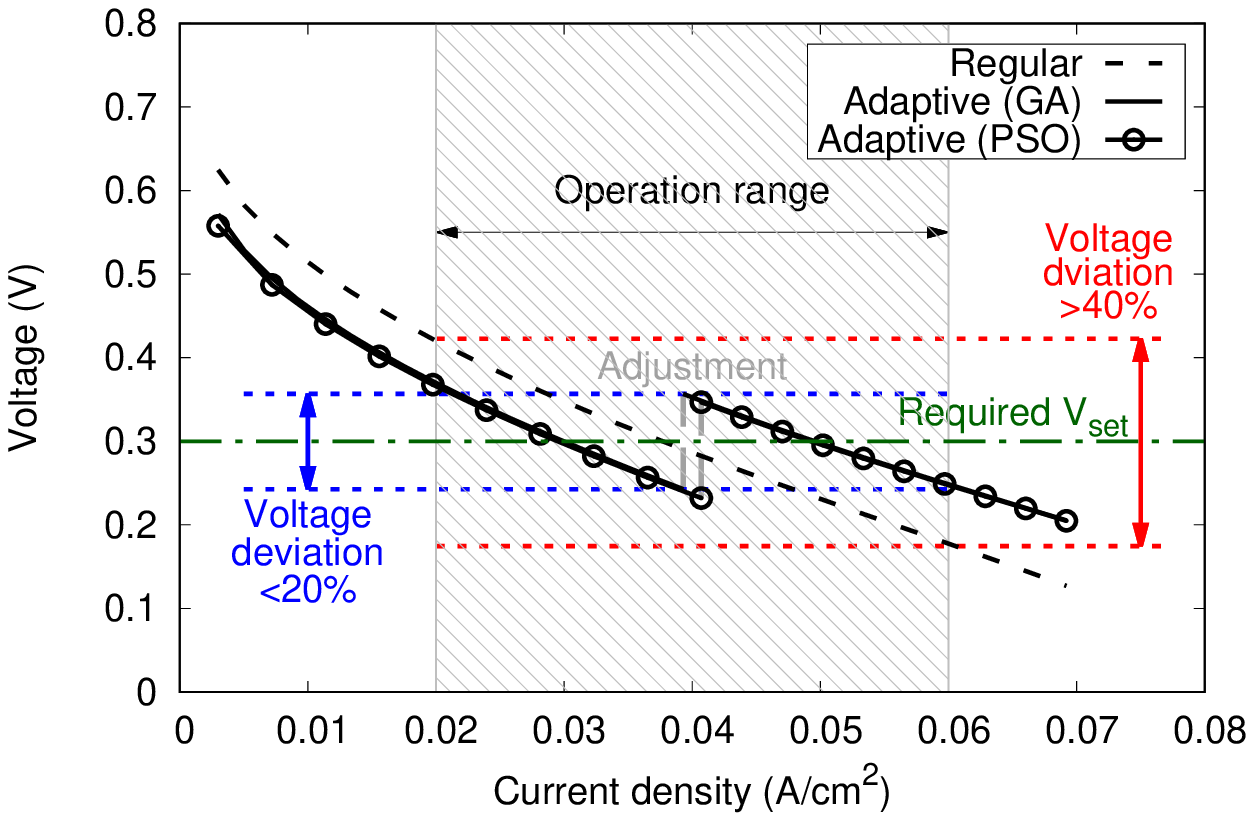}
\caption{Large voltage variation in a fixed range of current density (top) and the voltage stability enhancement using regular or adaptive operation strategy (bottom).}
\label{fig:4}
\end{figure}

Large voltage variation, that was revealed in many pioneering works \cite{Taner2018energy, Mehmood2015JPS, zhou2018ECM}, has also been frequently encountered in our experiments.
Two typical cases are presented in Fig. \ref{fig:4} (top).
They were observed in unoptimized experimental process, for which the operating parameters are summarized in Table \ref{tab:testcase}. 
As the current density varies in a fixed range (from 0.02 to 0.06 $\rm A/cm^2$), the output voltage is found to decrease from 0.3556 V to 0.0768 V for Case 1, and from 0.4638 V to 0.2092 V for Case 2.
It implies that the overpotential effect cannot be negligible for the operational performances of DMFC systems, as the increase in current density is accompanied by a significant decrease of output voltage (which can be larger than 50\%).
Such a huge voltage variation is also normally beyond the input limit of voltage stabilizers, and the operation strategy is therefore in great need to minimize the voltage variation to a desired degree.

\subsection{Optimized operations}
\label{Sec:OPToperation}

Principal aim of the proposed operation strategy is to minimize the voltage deviation as the current density changes in a required range.
Assuming that the permitted output voltage $V_{set}$ equals to 0.3 V and the current density varies between 0.02 and 0.06 $\rm A/cm^2$, the adaptive operations proposed by both regular and adaptive methods have been carefully examined.
The operating condition that ensures the smallest voltage deviation can be determined through a regular optimization using genetic algorithm, which is $T=322.8~\rm K$, $C_M=0.6~\rm M$, $F_M=5.5~\rm ccm$ and $F_A=140.8~\rm ccm$. 
As shown in Fig. \ref{fig:4} (bottom), the voltage deviation can be deduced to about $\pm40\%$ (-41.8\% to 40.93\%) by regular optimizations.
Although it is much better than those unoptimized cases (Fig. \ref{fig:4} (top)), it is still far from the input requirements of regular voltage stabilizers. How to effectively reduce the voltage deviation is still significant for practical applications of fuel cell systems.

As shown in Fig. \ref{fig:4} (bottom), the voltage deviation can be effectively decreased to less than 20\% using the adaptive operation strategies either by genetic algorithm (GA) or by particle swarm optimization (PSO).
Compared with regular optimization that only provides one set of fixed operating parameters, the adaptive optimization proposes a real-time control of operating conditions.
For instance, an operational adjustment is imposed at around the current density of 0.04 $\rm A/cm^2$, and the adaptive I-V curve using GA is therefore composed by the gentle-slope regions of two different I-V curves.
The left I-V curve is implemented at $T=325.3~\rm K$, $C_M=1.2~\rm M$, $F_M=5.4~\rm ccm$ and $F_A=140.6~\rm ccm$, and the right one is at $T=342.3~\rm K$, $C_M=0.3~\rm M$, $F_M=5.5~\rm ccm$ and $F_A=140.7~\rm ccm$. 
As the connected I-V curves are shown to be able to significantly subside the voltage variation, this type of real-time control of operating parameters can be regarded as 'optimized flexible operations'.
The real-time analysis and control of fuel cell systems have been widely studied \cite{wu2017realtimeIEEE, zhou2018ECM}, which have provided sufficient evidence for the realizability and feasibility of the proposed 'optimized flexible operations'.
It can be also important to notice that the operation strategies optimized by GA or PSO almost coincide with each other. 
It preliminarily shows that the proposed strategy has not been restricted to a specific optimization method.
However, as the present study principally aims at the formulation of adaptive control strategy, the effects of different optimization methods on the strategy performance will not be fully discussed.
The classical optimization method of GA is then adopted in the following results.

\subsection{Numerical analysis}

\begin{figure}[t]
\centering
\includegraphics[width=0.975\linewidth]{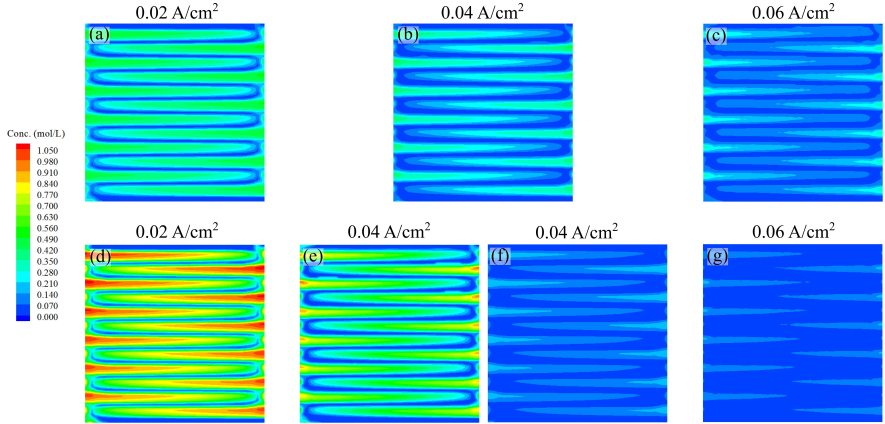}
\caption{Distribution of methanol concentration  on the interface plane of diffusion and catalyst layers. Top figures (a-c) represent regular operations, and the bottom ones (d-g) correspond to adaptive operations.}
\label{fig:NumResExplain}
\end{figure}

Numerical analysis is performed to answer the question "How does the adaptive operation strategy work on the voltage stability enhancement in DMFCs?".
Simulations are carried out using the constructed 3D numerical model \cite{30Hu2017Energies}. 
The operating conditions are chosen to be the same as that for Fig. \ref{fig:4} (bottom) in Section \ref{Sec:OPToperation}.

\begin{figure}[t]
\centering
\psfrag{Case 1}[c][c][0.675][0]{Case 1}
\psfrag{Case 2}[c][c][0.675][0]{Case 2}
\psfrag{Regular}[c][c][0.625][0]{Regular}
\psfrag{Adaptive}[c][c][0.625][0]{Adaptive}
\includegraphics[width=0.49\linewidth]{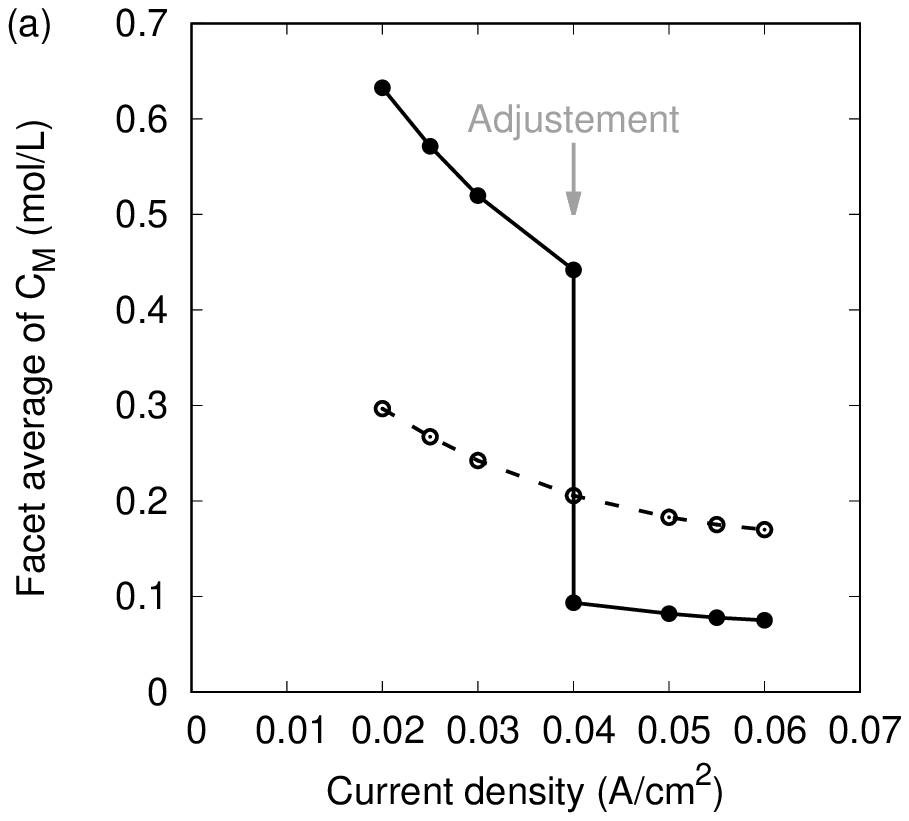}
\includegraphics[width=0.4875\linewidth]{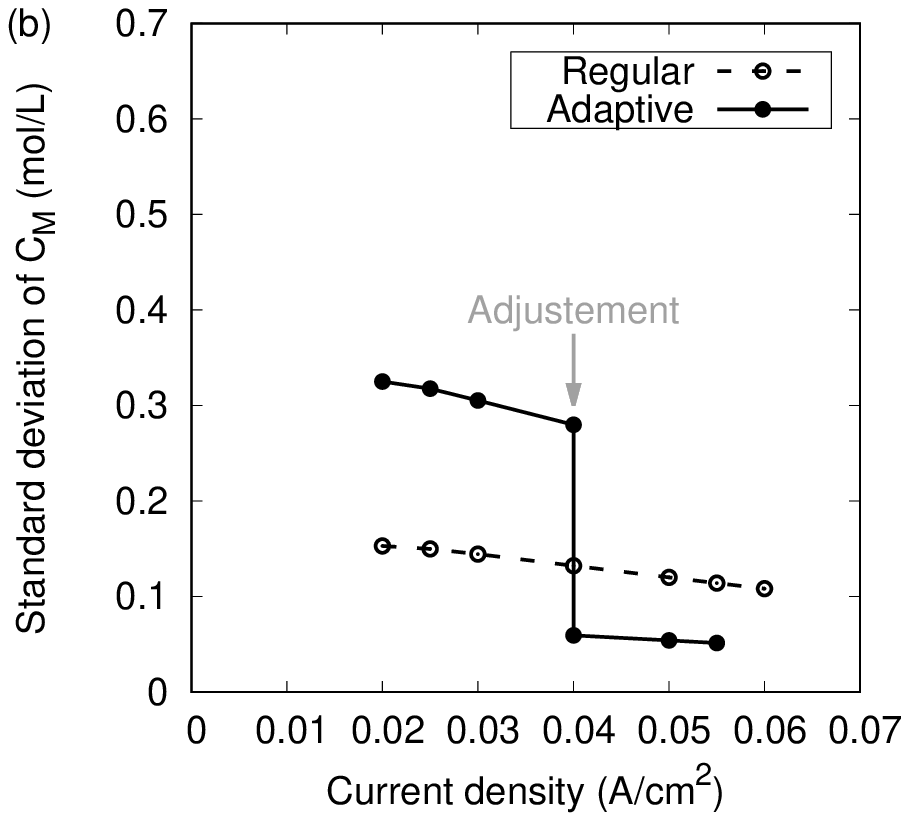}
\caption{Facet average (a) and standard deviation (b) of methanol concentration on the interface plane between the diffusion and catalyst layers.}
\label{fig:QuantFASD}
\end{figure}

Fig. \ref{fig:NumResExplain} shows the numerical results of methanol concentration distributing on the interface plane between the diffusion and catalyst layers, in which the top figures (a-c) represent regular operations and the bottom ones (d-g) represent adaptive operations.
A remarkable characteristics is the stripped distribution of methanol concentrations, which is mainly caused by the parallel serpentine arrangement of flow channel in DMFC systems \cite{26Yu2013IJHE, 30Hu2017Energies}.
It coincides very well with previous revealing, and more discussions about the effects of channel geometric patterns can be available in \cite{zhou2018ECM, Kianimanesh2013FC}.

For the regular case, the overall color temperature becomes increasingly low as the current density increases from 0.02 to 0.04 and to 0.06 $\rm A/cm^2$ (Fig. \ref{fig:NumResExplain} (a-c)).
It indicates that the methanol concentration decreases with the increase of current density, as the flux of methanol solution should be constant for a given operating condition. This result agrees with the physical fact that a larger power generation is normally based on a higher fuel consumption \cite{ZHOU2017IEEE, 26Yu2013IJHE, 30Hu2017Energies}.
A similar result can also be found in adaptive operations. 
However, as the adaptive operation strategy imposes an adjustment of operating parameters, there are two figures (e, f) at the current density 0.04 $\rm A/cm^2$ indicating the change of methanol concentration distributions before and after the adjustment.
The adaptive adjustment has caused a significant decrease of methanol concentration, which can be the reason for the sharp voltage increase in adaptive operations (as shown in Fig. \ref{fig:4} (bottom)).
At a low current density 0.02 $\rm A/cm^2$, the methanol concentration for the adaptive case (Fig. \ref{fig:NumResExplain} (d)) is much higher than that with regular optimization (Fig. \ref{fig:NumResExplain} (a)), but the situation reverses when a high current density 0.06 $\rm A/cm^2$ is reached.

Quantitative analysis on the methanol concentration distributing on the interface plane between diffusion and catalyst layers is summarized in Fig. \ref{fig:QuantFASD}.
As the current density varies from 0.02 to 0.06 $\rm A/cm^2$, the facet average of methanol concentration gradually decreases from 0.3 to 0.17 mol/L for regular case, while it declines sharply from 0.63 to 0.08 mol/L for the adaptive one (Fig. \ref{fig:QuantFASD} (a)). 
A slump occurring at 0.04 $\rm A/cm^2$ for adaptive case is caused by the operational adjustment that has also been demonstrated in Fig. \ref{fig:4} (bottom) and Fig. \ref{fig:NumResExplain}.
It is of importance to notice that, the slope of facet average of methanol concentration is significantly modified by the adjustment. 
Compared with regular optimization, it decreases faster for adaptive one before the operational adjustment. However, 
a slower decrease is achieved after the operational adjustment. 
It suggests that the adaptive operation strategy can effectively modulate the methanol consumption rate inside DMFCs to improve the energy conversion process. 
The regulation effect of adaptive operation strategy can also be found in results of standard deviation of methanol concentration.
As shown in Fig. \ref{fig:QuantFASD} (b), the uniformity of methanol concentration distributions can be markedly regulated by the operational adjustment. 
The above results have shown, in both qualitative and quantitative ways, that the operational adjustment proposed by adaptive operation strategy is an effective way to regulate the methanol consumption.
It can be highly beneficial for the maintenance of voltage stability in DMFCs that are required to be operated in a wide range of current density.

\subsection{On-demand adaptive operations}

\begin{figure}[t]
\centering
\includegraphics[width=0.725\linewidth]{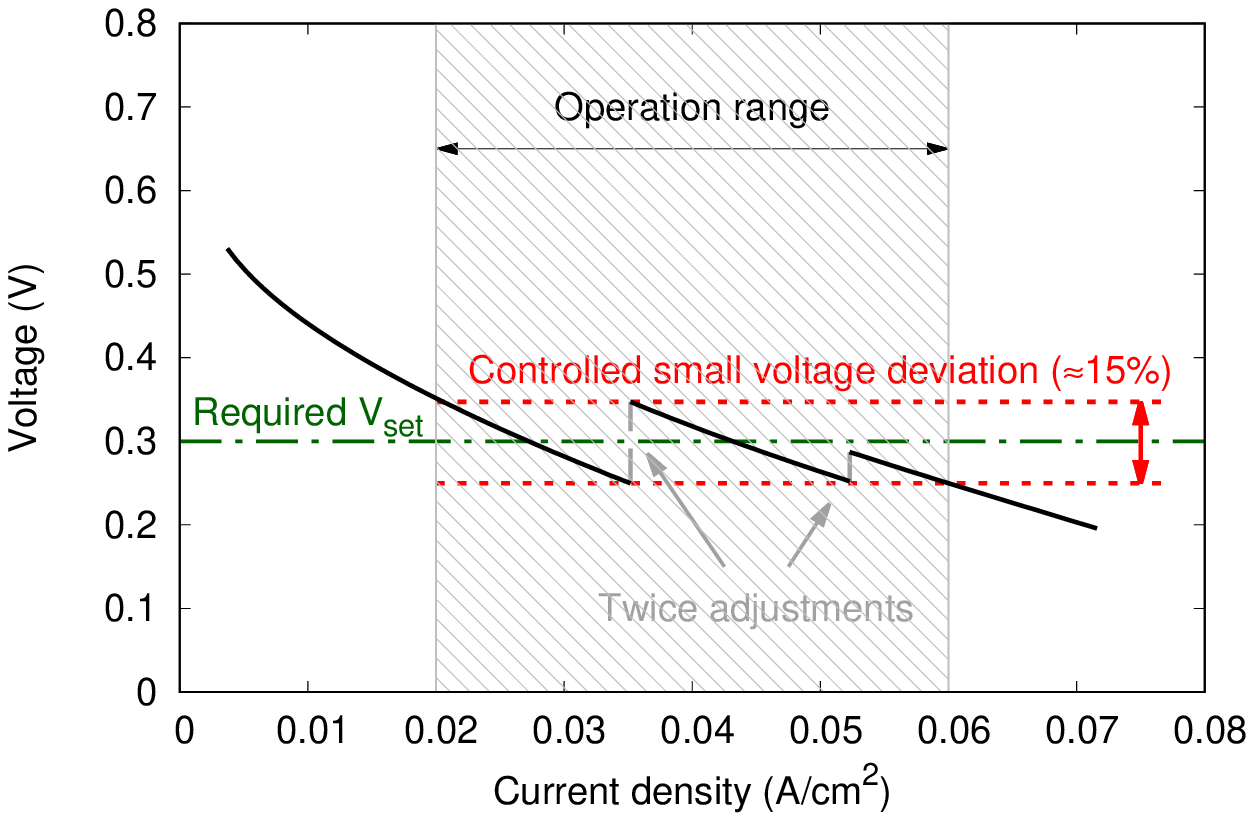}
\includegraphics[width=0.725\linewidth]{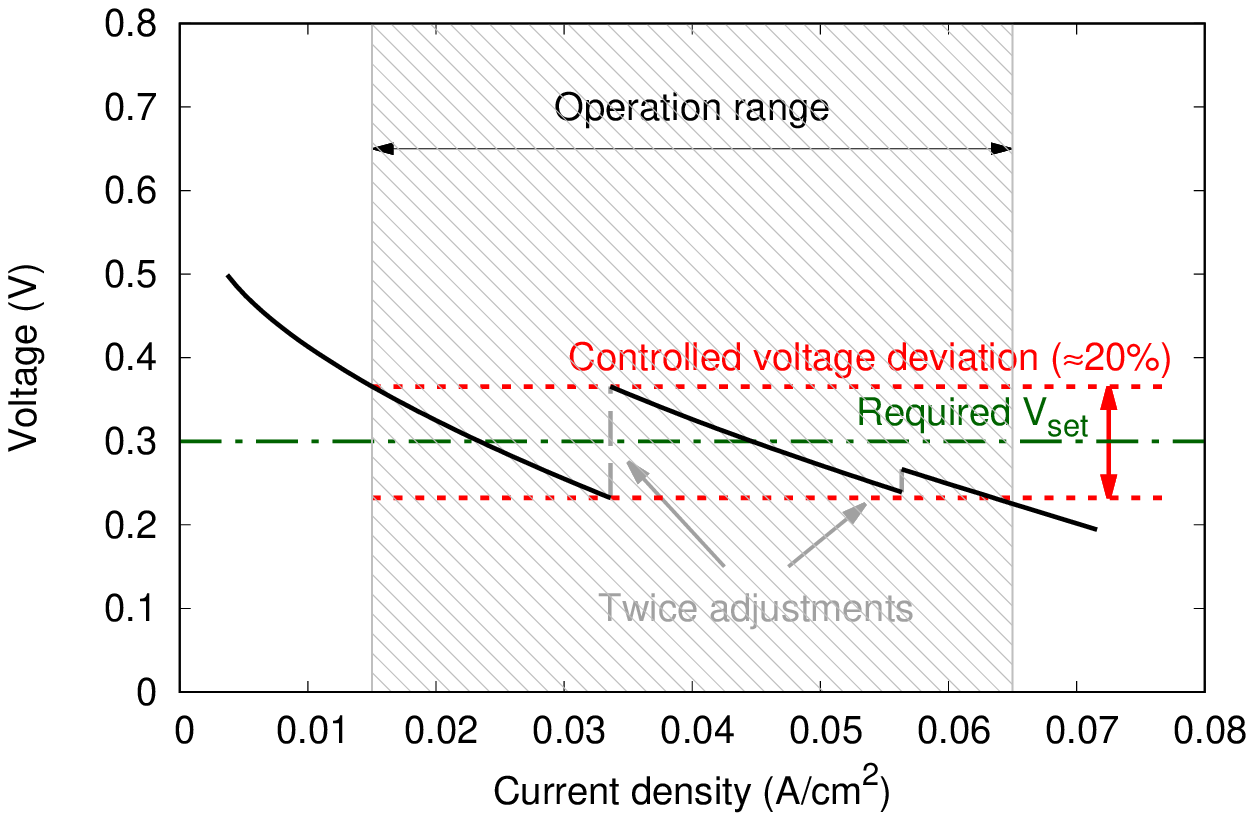}
\caption{On-demand control of voltage stability using adaptive operation strategy.}
\label{fig:adaptopt}
\end{figure}

Flexible operations are usually demanded in practical use, for example, decreasing the voltage deviation or extending the range of operating current density.
Such a flexible requirement can also be satisfied using on-demand control of operating parameters based on the adaptive operation strategy.
Compared with the constant requirements realized by optimized operations in Section \ref{Sec:OPToperation} (Fig. \ref{fig:4} (bottom), voltage deviation $\approx 20\%$ and current density varying in [0.02, 0.06]), the flexible operations with a smaller voltage deviation or a larger range of operating current density are considered in this section.

Fig. \ref{fig:adaptopt} shows the optimized adaptive operations according to user's requirements. For a required output voltage 0.3 V, the voltage deviation is successfully decreased to about $\pm$15\% in the first plot, while the operating current density is extended to [0.015, 0.065] $\rm A/cm^2$ in the second one. 
The cost for those high requirements is the increase of adjusting times. 
As shown in Fig. \ref{fig:adaptopt}, one needs to implement twice adjustments to achieve a more stable voltage output in a larger working current density. 
Taking the first plot as example, the left I-V curve section in [0.02, 0.035] $\rm A/cm^2$ corresponds to the operation conditions of $T=314.7~\rm K$, $C_M=1.1~\rm M$, $F_M=5.5~\rm ccm$ and $F_A=136.3~\rm ccm$, the middle one in [0.035, 0.052] $\rm A/cm^2$ corresponds to the operating conditions of $T=339.9~\rm K$, $C_M=0.3~\rm M$, $F_M=5.5~\rm ccm$ and $F_A=126.8~\rm ccm$, and the right one in [0.05, 0.06] $\rm A/cm^2$ corresponds to the operating conditions of $T=342.8~\rm K$, $C_M=0.3~\rm M$, $F_M=5.5~\rm ccm$ and $F_A=140.6~\rm ccm$.
All of them are determined by the adaptive operation strategy which ensures the smallest slope of each curve section in the corresponding interval of current density.

It is known to all that the DMFC system can provide a relatively stable voltage when it is operated in the gentle-slope range in I-V curves \cite{Mehmood2015JPS, Barbir2012PEM, 30Hu2017Energies}.
The adaptive operation strategy provides us a method to search and to integrate gentle-slope sections in different I-V curves (which takes a low rate of voltage decrease as current density increases), to ensure the required voltage stability and to meet various requirements from DMFC users.

\section{Experimental validations}
\label{Sec:ExpValid}

\subsection{Semi-empirical validation}

\begin{figure}[htpb]
\centering
\psfrag{Case 1}[r][r][0.75][0]{Case 1}
\psfrag{Case 2}[r][r][0.75][0]{Case 2}
\psfrag{Case 3}[r][r][0.75][0]{Case 3}
\includegraphics[width=0.75\linewidth]{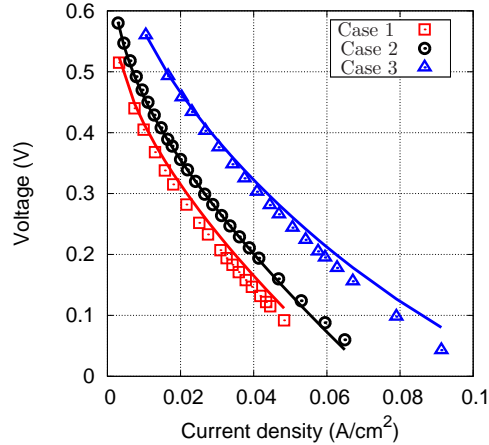}
\caption{Comparisons of experimental I-V data and corresponding predications using semi-empirical model (Line: Model; Dot: Experiment.)}
\label{fig:3ValidSemimodel}
\end{figure}

A simplified semi-empirical model has been constructed for predicating I-V relationships of DMFC system in the present study (as described in Section \ref{Sec:semi-empiriMod}).
Experimental tests, based on the experimental platform described in Section \ref{ExpSetup}, are performed for model validations. The operating parameters are summarized in Table \ref{tab:OperatSemiModel}.
Comparisons of experimental I-V data and semi-empirical predications are shown in Fig. \ref{fig:3ValidSemimodel}.
The experimental results can be well estimated by the proposed semi-empirical model, even if the deviation from model predications to experimental data would slightly increase as the operating voltage is approaching 0 V.
As the DMFC performance normally decays at a very high rate while its current density is approaching to its maximum values (i.e., output voltage approaches 0 V), the preferable operations normally happen at the intermediary regions of I-V curves.
As those 'extreme' DMFC operations are out of the scope of the present study, the good agreement of model predications to experimental results in Fig. \ref{fig:3ValidSemimodel} has validated the feasibility and effectiveness of the proposed semi-empirical model.

\begin{table}[t]
\centering
\caption{Operating parameters of testing cases for semi-empirical model validations.}
\label{tab:OperatSemiModel}
\begin{tabular}{>{\small}c >{\small}c >{\small}c >{\small}c >{\small}c}
\hline
$Case$ & $T$ (\rm{K}) & $C_M (\rm{M})$ & $F_M$ (\rm{ccm}) & $F_A$ (\rm{ccm}) \\
\hline
1 & 298 & 1.00 & 4.5 & 125.2 \\
2 & 323 & 0.25 & 4 & 81.2 \\
3 & 343 & 0.50 & 5 & 140.8 \\
\hline
\end{tabular}
\end{table}

\subsection{Controller validation}
\label{Sec:ContValid}

Experimental work has also been carried out to validate the effects of adaptive operation strategy on voltage stability enhancement.
It is based on a customized DMFC stack that integrates a  $\rm 5 \times 5 ~cm^2$ MEA with Nafion 115 membrane, in which Pt-Ru black (4 $\rm  mg/cm^2$) and Pt-C (0.3 $\rm  mg/cm^2$) are used as anode and cathode catalysts. 
The DMFC stack is required to work at an output voltage 0.38 V, and the current density lies in the range of [0.5, 2.5] $\rm A/cm^2$.

\begin{figure}[t]
\centering
\psfrag{Unoptimized}[c][c][0.675][0]{Unoptimized}
\psfrag{Adaptive OPT}[c][c][0.675][0]{Adaptive OPT}
\psfrag{Single parameter}[c][c][0.675][0]{Single parameter}
\psfrag{Double parameters}[c][c][0.675][0]{Double parameters}
\includegraphics[width=0.715\linewidth]{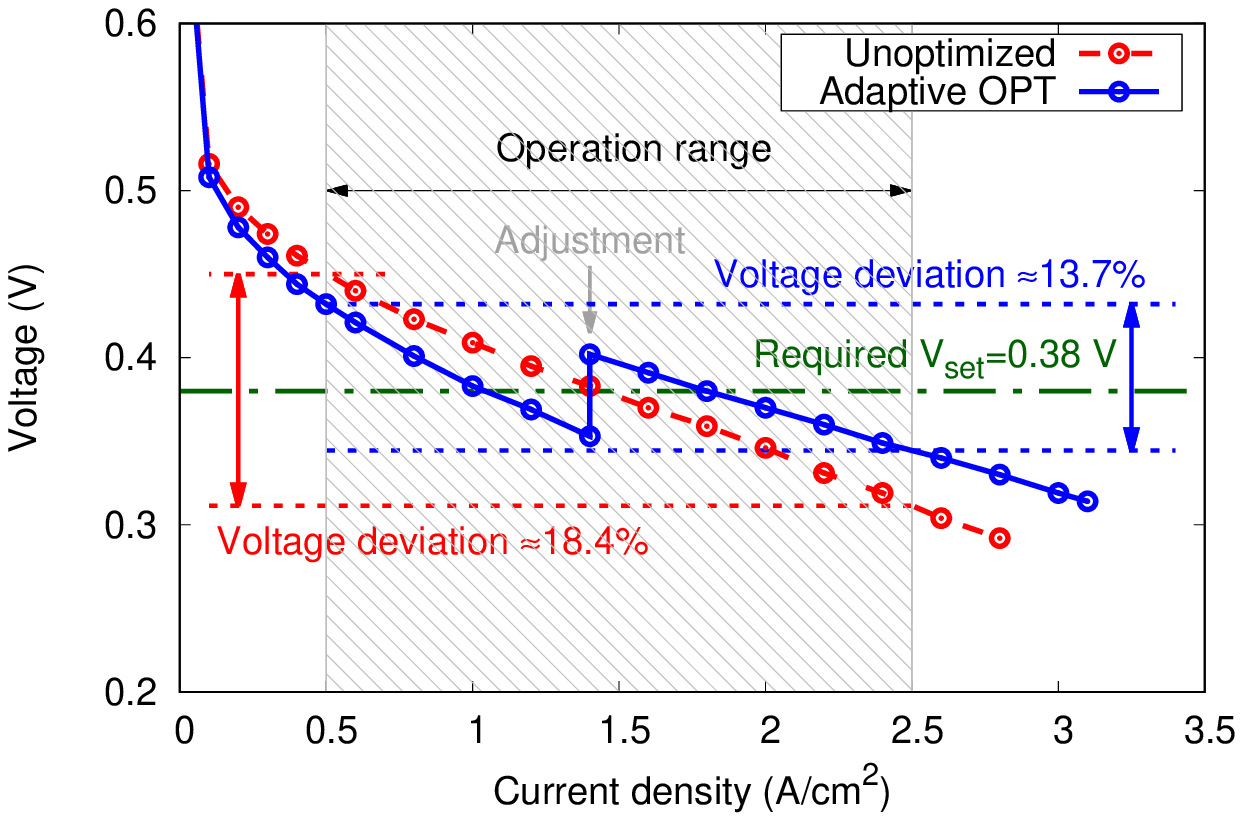}
\includegraphics[width=0.715\linewidth]{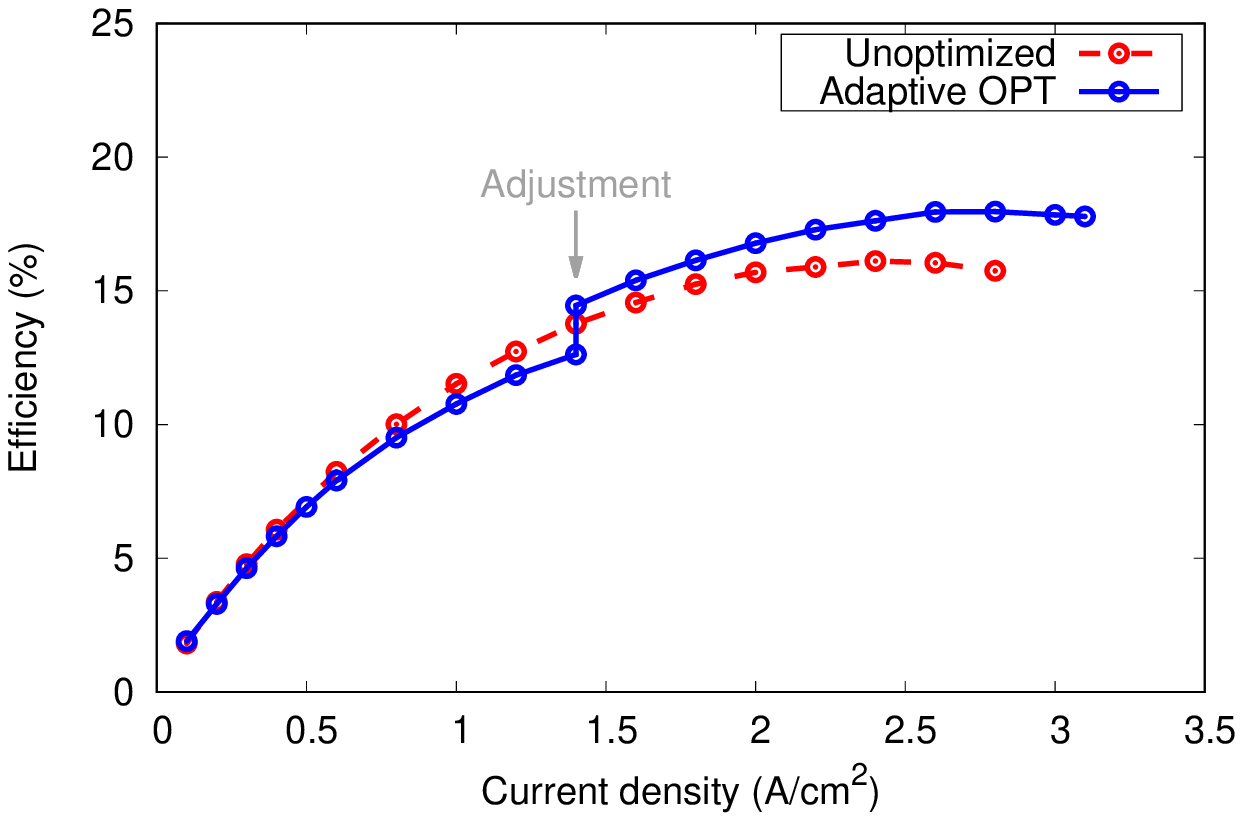}
\caption{Experimental validation of adaptive optimized operations of a customized DMFC system (top) and the corresponding efficiency analysis (bottom).}
\label{fig:OPTexp}
\end{figure}

Fig. \ref{fig:OPTexp} (top) shows the comparisons of experimental I-V curves obtained with or without adaptive operation strategy. 
The voltage deviation is significantly decreased from 18.4\% to 13.7\% as the adaptive operation is implemented. 
It implies that the voltage stability can be effectively enhanced by the adaptive operation strategy.
Compared with the unoptimized operation with a specific series of operation parameters of $T=338.0~\rm K$, $C_M=0.5~\rm M$, $F_M=2.0~\rm ccm$ and $F_A=1000.0~\rm ccm$, a manual adjustment of operating parameters should be imposed at 1.4 $\rm A/cm^2$ for adaptive operation. 

The efficiency analysis has also been performed to further investigate the effects of adaptive operation strategy on DMFC systems.
Several pioneering methods were successfully proposed for cost evaluations in various energy conversion systems \cite{Taner2015APE, silva2005JPS, 16casalegno2011JPS, Taner2017RSER}.
We apply one of the classical evaluation criteria that has been widely used in DMFC systems. It takes a form as follows,
\begin{align}
\centering 
\eta = \frac{V_{cell}I}{LHV (N_{cons} + N_{cross})},
\end{align}
where $\eta$ denotes the energy conversion efficiency, $V_{cell}$ and $I$ are the output voltage and current density, respectively. $LHV$ is the lower heating value of methanol, $N_{cons}$ and $N_{cross}$ denote the molar flow rates of methanol for effective current generation and the methanol crossover, respectively. 
It measures the energy conversion efficiency by the ratio of effective fuel cost, i.e., the percentage of effectively consumed fuel in the total amount of fuel 'investment'.

Fig. \ref{fig:OPTexp} (bottom) shows the energy conversion efficiency obtained in DMFC systems with or without adaptive operation strategy, which correspond to the cases in Fig. \ref{fig:OPTexp} (top). 
For low current density $\le 0.5~\rm A/cm^2$, the efficiencies of different operations are almost the same.
However, the difference becomes increasingly obvious as the operating current density increases.
An obvious jump can be found in adaptive operation, which is caused by the operational adjustment proposed by adaptive operation strategy. 
Actually, the mean efficiency of adaptive operation (12.2\%) is slightly higher than that of unadjusted one (11.4\%). 
It implies that the adaptive operation strategy can provide possibilities for voltage stability enhancement without the sacrifice of energy conversion efficiency.

\subsection{System response to adjustments}
\begin{figure}[t]
\centering
\psfrag{Unoptimized}[c][c][0.675][0]{Unoptimized}
\psfrag{Adaptive OPT}[c][c][0.675][0]{Adaptive OPT}
\psfrag{Single parameter}[c][c][0.675][0]{Single parameter}
\psfrag{Double parameters}[c][c][0.675][0]{Double parameters}
\includegraphics[width=0.725\linewidth]{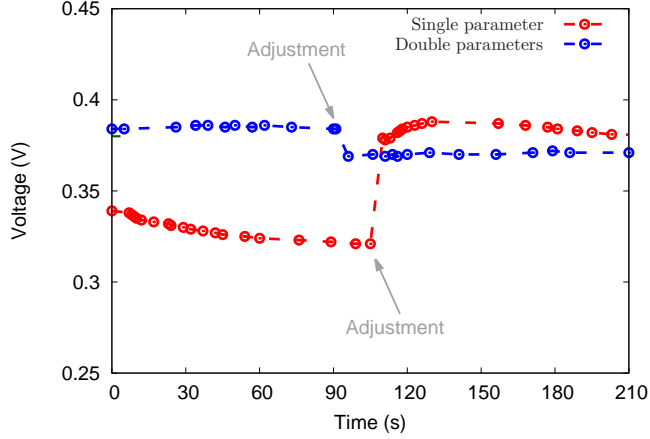}
\caption{System response to the adjustment of operating parameters.}
\label{fig:SystemResp}
\end{figure}
Although the operational adjustments in adaptive operation strategy is helpful for voltage stability enhancement, it can be also regarded as system disturbances that would bring out negative effects to system stability. 
The response of DMFC system to operational adjustments, therefore, needs to be taken into account in one's evaluations of the effectiveness of the adaptive operation strategy. 
Preliminary study has been carried out based on the customized DMFC stack (as described in Section \ref{Sec:ContValid}). 

Fig. \ref{fig:SystemResp} shows the time evolution of output voltage in a DMFC system, which is modulated by the adjustments of one or two operation parameters.
For single-parameter adjustment, the input air flow rate $F_A$ is increased from 200 to 800 $\rm ccm$, while the other operating parameters are kept at $T=328.0~\rm K$, $C_M=0.5~\rm M$, $F_M=3.0~\rm ccm$.
It can be found that the output voltage is relatively steady at about 0.34 $\rm V$ before the operational adjustment is imposed. 
After the single-parameter adjustment, a quick response takes place in the DMFC system, which results in an obvious voltage increase from 0.34 to 0.38 $\rm V$ in 5 seconds.
The same phenomenon has also been revealed for double-parameter adjustment, in which the input air flow rate $F_A$ is decreased from 500 to 100 $\rm ccm$, the methanol solution flow rate $F_M$ is decreased from 3.0 to 1.0 $\rm ccm$, and the other two operating parameters are kept at $T=328.0~\rm K$, $C_M=0.5~\rm M$.

It shows that the response time of DMFC system to operational adjustments can be as short as less than 5 seconds.
Such a rapid response ensures, to some extent, the realizability of the proposed adaptive operation strategy.
However, it is well-known that, the flow rates of the air and the feed solution are relatively easy to be modulated by manual operations. 
In order to realize an automatic adaptive operation, an automatic operating system embedded with adaptive operation strategy can be highly beneficial for the adjustment implementations. 
It will be considered in our future work.

\section{Conclusions}
\label{Sec:ConPers}

Concentrating on the problem of voltage instability, an adaptive operation strategy was successfully developed for on-demand control of DMFC systems to enhance the voltage stability in operating DMFC systems.
The procedure of strategy developing has been discussed in details in the present study, and the feasibility and superiority of the proposed operation strategy have been systematically studied using both experimental and numerical techniques.
The main findings in the present study can be summarized as follows,

1) Large voltage variation was frequently observed in the experiments using a typical DMFC stack.
Although it can be decreased by a regular optimization method to a certain degree, it still needs improvements to avoid additional energy loss caused by the process of voltage stabilization.
The adaptive operation strategy can be a potential solution, as it could effectively decrease the voltage variation from more than 40\% to less than 20\% by introducing certain operational adjustments.

2) On-demand DMFC operations, such as providing a smaller voltage deviation or a larger operating current density, was shown to be fulfilled by the adaptive operation strategy. 
The adaptive operation strategy provides a method to search and to integrate gentle-slope sections in different I-V curves to ensure the demanded voltage stability in flexible operations.

3) Numerical results were provided to study the underlying mechanisms of the adaptive operation strategy, as the adjustment of operating parameters proposed by adaptive operation strategy can be an effective way to regulate the methanol consumption inside the DMFC system.
Experimental work was also carried out to validate the effects of adaptive operation strategy on the voltage stability in an operating DMFC system. 
A rapid response of DMFC system (less than 5 seconds) to the change of operating parameters was experimentally observed. It ensures the applicability of the proposed adaptive operation strategy.

\section*{Acknowledgments}
Project supported by the National Natural Science Foundation of China (No.\,51505136 \& 11402084), the Natural Science Foundation of Hunan Province (No. 2015JJ3051 \& 2017JJ3018), the self-determined project of State Key Laboratory of Advanced Design and Manufacturing for Vehicle Body (No. 51475002), the open fund of State Key Laboratory of Fluid Power \& Mechatronics Systems (No.\, GZKF-201615) and the Fundamental Research Funds for the Central Universities (Hunan University).

\section*{References}
\bibliography{Ref_ECM}

\begin{thebibliography}{10}
\expandafter\ifx\csname url\endcsname\relax
  \def\url#1{\texttt{#1}}\fi
\expandafter\ifx\csname urlprefix\endcsname\relax\def\urlprefix{URL }\fi
\expandafter\ifx\csname href\endcsname\relax
  \def\href#1#2{#2} \def\path#1{#1}\fi

\bibitem{Chau1999ECM}
K.~Chau, Y.~Wong, C.~Chan, An overview of energy sources for electric vehicles,
  Energy Conversion and Management 40~(10) (1999) 1021--1039.

\bibitem{Ji2017ECM}
F.~Ji, L.~Yang, H.~Sun, S.~Wang, H.~Li, L.~Jiang, G.~Sun, A novel method for
  analysis and prediction of methanol mass transfer in direct methanol fuel
  cell, Energy Conversion and Management 154 (2017) 482--490.

\bibitem{Kamaruddin2013RSER}
M.~Kamaruddin, S.~Kamarudin, W.~Daud, M.~Masdar, An overview of fuel management
  in direct methanol fuel cells, Renewable and Sustainable Energy Reviews 24
  (2013) 557--565.

\bibitem{Taner2018energy}
T.~Taner, Energy and exergy analyze of pem fuel cell: A case study of modeling
  and simulations, Energy 143 (2018) 284--294.

\bibitem{Taner2015APE}
T.~Taner, Optimisation processes of energy efficiency for a drying plant: {A}
  case of study for turkey, Applied Thermal Engineering 80 (2015) 247--260.

\bibitem{Zhou2016IEEE}
D.~Zhou, E.~Breaz, A.~Ravey, F.~Gao, A.~Miraoui, K.~Zhang, Dynamic variable
  coupling analysis and modeling of proton exchange membrane fuel cells for
  water and thermal management, in: Applied Power Electronics Conference and
  Exposition (APEC), 2016 IEEE, IEEE, 2016, pp. 3476--3480.

\bibitem{ZHOU2017IEEE}
D.~Zhou, F.~Gao, E.~Breaz, A.~Ravey, A.~Miraoui, Tridiagonal matrix algorithm
  for real-timetime simulation of a 2-d pem fuel cell model, IEEE Transactions
  on Industrial Electronics PP~(99) (2017) 1--1.
\newblock \href {http://dx.doi.org/10.1109/TIE.2017.2787598}
  {\path{doi:10.1109/TIE.2017.2787598}}.

\bibitem{wu2017realtimeIEEE}
C.~Wu, J.~Chen, C.~Xu, Z.~Liu, Real-time adaptive control of a fuel
  cell/battery hybrid power system with guaranteed stability, IEEE Transactions
  on Control Systems Technology 25~(4) (2017) 1394--1405.

\bibitem{karim2013CatalystAE}
N.~Karim, S.~Kamarudin, An overview on non-platinum cathode catalysts for
  direct methanol fuel cell, Applied Energy 103 (2013) 212--220.

\bibitem{Patel2015JPS}
P.~Patel, M.~Datta, P.~Jampani, D.~Hong, J.~Poston, A.~Manivannan, P.~Kumta,
  High performance and durable nanostructured {T}i{N} supported
  $\rm{Pt}_{50}$--$\rm{Ru}_{50}$ anode catalyst for direct methanol fuel cell
  ({DMFC}), Journal of Power Sources 293 (2015) 437--446.

\bibitem{fukuhara2014SSI}
L.~Fukuhara, N.~Kado, K.~Kosugi, P.~Suksawad, Y.~Yamamoto, H.~Ishii,
  S.~Kawahara, Preparation of polymer electrolyte membrane with nanomatrix
  channel through sulfonation of natural rubber grafted with polystyrene, Solid
  State Ionics 268 (2014) 191--197.

\bibitem{Wang2017ECM}
M.~Wang, G.~Liu, Z.~Tian, Y.~Shao, L.~Wang, F.~Ye, M.~Tran, Y.~Yun, J.~Lee,
  Microstructure-modified proton exchange membranes for high-performance direct
  methanol fuel cells, Energy Conversion and Management 148 (2017) 753--758.

\bibitem{Ozden2017IJHE}
A.~Ozden, M.~Ercelik, D.~Ouellette, C.~Colpan, H.~Ganjehsarabi,
  F.~Hamdullahpur, Designing, modeling and performance investigation of
  bio-inspired flow field based {DMFC}s, International Journal of Hydrogen
  Energy (2017) 1--13.

\bibitem{Wilberforce2017IJHE}
T.~Wilberforce, Z.~El-Hassan, F.~Khatib, A.~Al~Makky, J.~Mooney, A.~Barouaji,
  J.~Carton, A.-G. Olabi, Development of {B}i-polar plate design of {PEM} fuel
  cell using {CFD} techniques, International Journal of Hydrogen Energy 42~(40)
  (2017) 25663--25685.

\bibitem{Mehmood2015JPS}
A.~Mehmood, M.~Scibioh, J.~Prabhuram, M.~An, H.~Ha, A review on durability
  issues and restoration techniques in long-term operations of direct methanol
  fuel cells, Journal of Power Sources 297 (2015) 224--241.

\bibitem{Taner2015}
T.~Taner, Alternative energy of the future: a technical note of {PEM} fuel cell
  water management, Journal of Fundamentals of Renewable Energy and
  Applications 5~(3) (2015) 1--4.

\bibitem{taner2017micro}
T.~Taner, The micro-scale modeling by experimental study in {PEM} fuel cell,
  Journal of Thermal Engineering 3~(6) (2017) 1515--1526.

\bibitem{Barbir2012PEM}
F.~Barbir, PEM fuel cells: theory and practice, Academic Press, 2012.

\bibitem{WANG2011reviewAE}
Y.~Wang, K.~Chen, J.~Mishler, S.~Cho, X.~Adroher, A review of polymer
  electrolyte membrane fuel cells: technology, applications, and needs on
  fundamental research, Applied Energy 88~(4) (2011) 981--1007.

\bibitem{Bizon2014ECM}
N.~Bizon, Load-following mode control of a standalone renewable/fuel cell
  hybrid power source, Energy Conversion and Management 77 (2014) 763--772.

\bibitem{Bizon2015ECM}
N.~Bizon, M.~Oproescu, M.~Raceanu, Efficient energy control strategies for a
  standalone renewable/fuel cell hybrid power source, Energy Conversion and
  Management 90 (2015) 93--110.

\bibitem{13Zenith2010JPC}
F.~Zenith, U.~Krewer, Modelling, dynamics and control of a portable {DMFC}
  system, Journal of Process Control 20~(5) (2010) 630--642.

\bibitem{16Chang2012IJICIC}
C.-Y. Chang, C.-H. Hsu, W.-J. Wang, C.-Y. Chen, The active control design for
  {DMFC/Battery} hybrid system using {PIDNN}, International Journal of
  Innovative Computing Information and Control 8~(3 B) (2012) 2101--2112.

\bibitem{17Fan2013JESTR}
L.~Fan, J.~Zhang, C.~Li, Model predictive control on constant voltage output of
  a proton exchange membrane fuel cell, Journal of Engineering Science and
  Technology Review 6~(2) (2013) 115--119.

\bibitem{18Keller2017CEP}
R.~Keller, S.~Ding, M.~M{\"{u}}ller, D.~Stolten, Fault-tolerant model
  predictive control of a direct methanol-fuel cell system with actuator
  faults, Control Engineering Practice 66 (2017) 99--115.

\bibitem{Zhou2017JPS}
D.~Zhou, A.~Al-Durra, F.~Gao, A.~Ravey, I.~Matraji, M.~Sim{\~o}es, Online
  energy management strategy of fuel cell hybrid electric vehicles based on
  data fusion approach, Journal of Power Sources 366 (2017) 278--291.

\bibitem{Zhou2017ECM}
D.~Zhou, A.~Ravey, A.~Al-Durra, F.~Gao, A comparative study of extremum seeking
  methods applied to online energy management strategy of fuel cell hybrid
  electric vehicles, Energy Conversion and Management 151 (2017) 778--790.

\bibitem{19Oliveira2008IJHE}
V.~Oliveira, D.~Falcao, C.~Rangel, A.~Pinto, Heat and mass transfer effects in
  a direct methanol fuel cell: A {1D} model, International Journal of Hydrogen
  Energy 33~(14) (2008) 3818--3828.

\bibitem{20Ko2010Energy}
J.~Ko, P.~Chippar, H.~Ju, A one-dimensional, two-phase model for direct
  methanol fuel cells--{P}art {I}: Model development and parametric study,
  Energy 35~(5) (2010) 2149--2159.

\bibitem{22Birgersson2003JES}
E.~Birgersson, J.~Nordlund, H.~Ekstr{\"o}m, M.~Vynnycky, G.~Lindbergh, Reduced
  two-dimensional one-phase model for analysis of the anode of a {DMFC},
  Journal of the Electrochemical Society 150~(10) (2003) A1368--A1376.

\bibitem{23Yan2008IJHMT}
T.~Yan, T.-C. Jen, Two-phase flow modeling of liquid-feed direct methanol fuel
  cell, International Journal of Heat and Mass Transfer 51~(5) (2008)
  1192--1204.

\bibitem{zhou2018ECM}
D.~Zhou, T.~Nguyen, E.~Breaz, D.~Zhao, S.~Cl{\'e}net, F.~Gao, Global parameters
  sensitivity analysis and development of a two-dimensional real-time model of
  proton-exchange-membrane fuel cells, Energy Conversion and Management 162
  (2018) 276--292.

\bibitem{26Yu2013IJHE}
B.~Yu, Q.~Yang, A.~Kianimanesh, T.~Freiheit, S.~Park, H.~Zhao, D.~Xue, A {CFD}
  model with semi-empirical electrochemical relationships to study the
  influence of geometric and operating parameters on {DMFC} performance,
  International Journal of Hydrogen Energy 38~(23) (2013) 9873--9885.

\bibitem{Heidary2016ECM}
H.~Heidary, M.~Kermani, B.~Dabir, Influences of bipolar plate channel blockages
  on {PEM} fuel cell performances, Energy Conversion and Management 124 (2016)
  51--60.

\bibitem{29Yang2011JPS}
Q.~Yang, A.~Kianimanesh, T.~Freiheit, S.~Park, D.~Xue, A semi-empirical model
  considering the influence of operating parameters on performance for a direct
  methanol fuel cell, Journal of Power Sources 196~(24) (2011) 10640--10651.

\bibitem{30Hu2017Energies}
X.~Hu, X.~Wang, J.~Chen, Q.~Yang, D.~Jin, X.~Qiu, Numerical investigations of
  the combined effects of flow rate and methanol concentration on {DMFC}
  performance, Energies 10~(8) (2017) 1094.

\bibitem{27Chu2006EA}
D.~Chu, R.~Jiang, Effect of operating conditions on energy efficiency for a
  small passive direct methanol fuel cell, Electrochimica Acta 51~(26) (2006)
  5829--5835.

\bibitem{28Silva2012AMC}
V.~Silva, A.~Rouboa, Optimizing the {DMFC} operating conditions using a
  response surface method, Applied Mathematics and Computation 218~(12) (2012)
  6733--6743.

\bibitem{zhou2016IEEEEC}
D.~Zhou, F.~Gao, E.~Breaz, A.~Ravey, A.~Miraoui, K.~Zhang, Dynamic phenomena
  coupling analysis and modeling of proton exchange membrane fuel cells, IEEE
  Transactions on Energy Conversion 31~(4) (2016) 1399--1412.

\bibitem{Fang2000Tech}
K.-T. Fang, D.~Lin, P.~Winker, Y.~Zhang, Uniform design: theory and
  application, Technometrics 42~(3) (2000) 237--248.

\bibitem{Kianimanesh2013FC}
A.~Kianimanesh, Q.~Yang, S.~Park, D.~Xue, T.~Freiheit, Model for the
  degradation performance of a single-cell direct methanol fuel cell under
  varying operational conditions, Fuel Cells 13~(6) (2013) 1005--1017.

\bibitem{asadi2014multi}
E.~Asadi, M.~da~Silva, C.~Antunes, L.~Dias, L.~Glicksman, Multi-objective
  optimization for building retrofit: {A} model using genetic algorithm and
  artificial neural network and an application, Energy and Buildings 81 (2014)
  444--456.

\bibitem{delgarm2016multi}
N.~Delgarm, B.~Sajadi, F.~Kowsary, S.~Delgarm, Multi-objective optimization of
  the building energy performance: {A} simulation-based approach by means of
  particle swarm optimization ({PSO}), Applied Energy 170 (2016) 293--303.

\bibitem{wang2015multi}
M.~Wang, J.~Wang, P.~Zhao, Y.~Dai, Multi-objective optimization of a combined
  cooling, heating and power system driven by solar energy, Energy Conversion
  and Management 89 (2015) 289--297.

\bibitem{marques2015multi}
J.~Marques, M.~Cunha, D.~Savi{\'c}, Multi-objective optimization of water
  distribution systems based on a real options approach, Environmental
  Modelling \& Software 63 (2015) 1--13.

\bibitem{silva2005JPS}
V.~Silva, S.~Weisshaar, R.~Reissner, B.~Ruffmann, S.~Vetter, A.~Mendes,
  L.~Madeira, S.~Nunes, Performance and efficiency of a {DMFC} using
  non-fluorinated composite membranes operating at low/medium temperatures,
  Journal of Power Sources 145~(2) (2005) 485--494.

\bibitem{16casalegno2011JPS}
A.~Casalegno, C.~Santoro, F.~Rinaldi, R.~Marchesi, Low methanol crossover and
  high efficiency direct methanol fuel cell: the influence of diffusion layers,
  Journal of Power Sources 196~(5) (2011) 2669--2675.

\bibitem{Taner2017RSER}
T.~Taner, M.~Sivrioglu, A techno-economic \& cost analysis of a turbine power
  plant: {A} case study for sugar plant, Renewable and Sustainable Energy
  Reviews 78 (2017) 722--730.

\end{thebibliography}
\end{spacing} 

\appendix
\section{Additional Data}

\begin{table}[htpb]
\centering
\caption{Operating parameters for uniform-designed experiments.}
\label{tab:expconditions}
\centerline{
\begin{tabular}{>{\small}c >{\small}c >{\small}c >{\small}c >{\small}c>{\small}c >{\small}c >{\small}c >{\small}c >{\small}c}
\hline
$ No. $ & $T ($\rm{K}$) $ & $C_M ($\rm{M}$) $ & $F_M$ ($\rm{ccm}$) & $F_A$ ($\rm{ccm}$) & $ No. $ & $T ($\rm{K}$) $ & $C_M ($\rm{M}$) $ & $F_M$ ($\rm{ccm}$) & $F_A$ ($\rm{ccm}$) \\
\hline
1 & 298 & 1    & 5.5 & 93.6   & 2    & 333 & 1      & 4.5 & 125.5   \\
3 & 323 & 2    & 4    & 81.2    & 4   & 333 & 0.25 & 3.5 &  81.2    \\
5 & 323 & 1.5 & 5    & 93.6    & 6   & 298 & 1.5   & 4    &  81.2    \\
7 & 323 & 0.5 & 4    & 125.2  & 8   & 323 & 0.5   & 4.5 &  125.2  \\
9 & 343 & 2    & 3.5 & 93.6    & 10 & 313 & 1      & 4    &  108.7 \\
11 & 298  & 0.5  & 4.5 & 125.2  & 12 & 343 & 1.5 & 4.5 & 81.2   \\
13 & 333  & 0.5  & 5.5 & 108.7  & 14 & 313 & 1.5 & 4.5 & 108.7 \\
15 & 323  & 1.5  & 5    & 125.2  & 16 & 323 & 0.25 & 5.5 & 93.6 \\
17 & 313  & 2     & 4.5 & 140.8  & 18 & 333 & 1.5   & 3.5 & 108.7 \\
19 & 313  & 1     & 3.5 & 81.2    & 20 & 323 & 1.5   & 3.5 & 140.8 \\
21 & 313  & 2     & 5.5 & 81.2    & 22 & 298 & 0.5   & 3.5 & 93.6 \\
23 & 313  & 0.25 & 3.5 & 108.7 & 24 & 298 & 1.5   & 5.5 & 140.8 \\
25 & 298  & 2      & 3.5 & 125.2 & 26 & 313 & 0.25 & 5.5 & 125.2 \\
27 & 333  & 1      & 5    & 81.2   & 28 & 313 & 0.5   & 4.5 & 93.6 \\
29 & 343  & 0.5   & 5.5 & 81.2   & 30 & 333 & 1      & 5.5 & 140.8 \\
31 & 343  & 0.25 & 5    & 140.8 & 32 & 333 & 2      & 4    &  140.8 \\
33 & 343  & 0.5   & 3.5 & 140.8 & 34 & 333 & 0.25 & 4.5 & 93.6 \\
35 & 343  & 1.5   & 5    & 108.7 & 36 & 298 & 0.25 & 5    & 81.2 \\
37 & 323  & 1      & 4    & 93.6   & 38 & 333 & 2      & 4.5 & 93.6 \\
39 & 298  & 0.25 & 4    & 140.8 & 40 & 343 & 0.25 & 4    & 108.7 \\
41 & 343  & 2      & 5.5 & 125.2 & 42 & 313 & 1      & 5    & 140.8 \\
43 & 298  & 2      & 5    & 108.7 & 44 & 323 & 0.5   & 5    & 108.7 \\
45 & 343  & 1      & 4    & 125.2 \\
\hline
\end{tabular} }
\end{table}

\end{document}